\documentstyle[11pt]{article}
\textwidth\def\kon#1#2{\vbox{\halign{##&&##\cr\lower4pt
\hbox{$\scriptscriptstyle\vert$}\hrulefill &\hrulefill\lower4pt
\hbox{$\scriptscriptstyle\vert$}\cr $#1$&$#2$\cr}}}

\def\al{\alpha}
\def\be{\beta}
\def\ga{\gamma}
\def\ro{\varrho}

\def\eh{{\scriptstyle{1\over 2}}}
\def\d{\partial}
\def\=d{\,{\buildrel\rm def\over =}\,}

\def\sqr#1#2{{\vcenter{\vbox{\hrule height.#2pt\hbox{\vrule width.
#2pt height#1pt \kern#1pt \vrule width.#2pt}\hrule height.#2pt}}}}

\def\sq{\hbox{\rlap{$\sqcap$}$\sqcup$}}

\def\Hp{{\cal H_{\rm phys}}}
\def\H{{\cal H}}

\begin{document}

\title{MASSIVE GRAVITY FROM DESCENT EQUATIONS}
\author{D.R. Grigore 
\footnote{e-mail: grigore@theor1.theory.nipne.ro, grigore@theory.nipne.ro}
\\ Department of Theoretical Physics, Inst. for Physics and Nuclear
Engineering "Horia Hulubei"
\\ Institute of Atomic Physics
\\Bucharest-Magurele, P.O.Box MG6, ROMANIA
\\ G. Scharf
\footnote{e-mail: scharf@physik.unizh.ch}
\\ Institut f\"ur Theoretische Physik, 
\\ Universit\"at Z\"urich, 
\\ Winterthurerstr. 190 , CH-8057 Z\"urich, SWITZERLAND}

\date{}

\maketitle\vskip 3cm

\begin{abstract} Both massless and massive gravity are derived
from descent equations (Wess-Zumino consistency conditions).
The massive theory is a continuous deformation of the massless one.

\end{abstract}

\newpage

\section{Introduction}

Gravity with massive gravitons has attracted much interest in recent
years in view of the evidence for dark energy. There exist very 
different approaches to
construct such a theory. We will follow a conservative route in
obtaining massive quantum gravity as a spin-2 gauge theory.
Indeed this approach was already realized successfully in a previous
work [1]. In this paper we have only used the lowest condition of gauge
invariance (equation (1.1) below). As a consequence the computations
have become rather involved. Here we work with a more complete gauge
structure expressed by the descent equations which will enable us to 
find the gravitational couplings in
the most elegant way and we shall get some modification of our previous
results. 

In our setting gauge theories are defined in a Fock space $\H$ with an
indefinite bilinear form, generated by physical and unphysical free
quantum fields. The physical states are selected by means of
the nilpotent gauge charge operator $Q$ ($Q^2=0$) such that the physical
Hilbert space is given by $\Hp={\rm Ker} Q/{\rm Ran}Q$. The space
$\H$ is endowed with a grading called ghost number and by
construction the gauge charge is raising the ghost number of a state
by one. Moreover, the space of Wick monomials of free fields in $\H$
is also endowed with a grading which follows by assigning a ghost
number to everyone of the free fields. The graded commutator $d_Q$ of
the gauge charge with any operator $A$ of fixed ghost number
$$d_QA=[Q,A]$$
is raising the ghost number by one. It means that $d_Q$ is a co-chain
operator in the space of Wick polynomials.

A gauge theory is given by a Wick polynomial of zero ghost number $T(x)$
called the interaction Lagrangian which satisfies
$$[Q,T(x)]=i\d_\al T^\al(x),\eqno(1.1)$$
where $T^\al$ is a Wick polynomial with ghost number 1. Since $d_Q$ and
the space-time derivative $\d_\al$ commute it follows from nilpotency that
$$\d_\al d_Q T^\al=0.$$
If the appropriate form of the Poincar\'e lemma is true, this implies
$$d_QT^\al=[Q,T^\al]=i\d_\be T^{\al\be}\eqno(1.2)$$
with antisymmetric $T^{\al\be}$. In the same way we get
$$[Q,T^{\al\be}]=i\d_\ga T^{\al\be\ga}\ldots\eqno(1.3)$$
with totally antisymmetric $T^{\al\be\ga}$ and so on. These are the
so-called descent equations (similar to Wess-Zumino consistency conditions).

However there are two obstacles for applying the usual Poincar\'e
lemma: first our co-cycles are polynomials and second we are working on
the mass shell; all fields obbey the Klein-Gordon equation with mass
$m$. If only the first obstacle would be present then we could apply the
so-called algebraic Poincar\'e lemma [2], but unfortunately this nice
result breaks down if we work on shell. As a consequence the above
argument does not prove the descent equations. It is a highly
non-trivial fact that, nevertheless, the descent equations hold in the
on-shell formalism. In the spin-2 case the chain of equations stopps
after (1.3). Working backwards from (1.3) to (1.1) is the descent
procedure. We will carry it out for the massless case $m=0$ in the next
section and for massive gravity in section 3.

\section{Massless quantum gravity}

The basic free asymptotic fields of massless gravity are the symmetric
tensor field $h^{\mu\nu}(x)$ with arbitrary trace and the fermionic ghost
$u^\mu(x)$ and anti-ghost $\tilde u^\mu(x)$ fields.
They all satisfy the wave equation
$$\sq h^{\mu\nu}=0=\sq u^\mu=\sq\tilde u    \eqno(2.1)$$
and are quantized as follows [3] [4]
$$[h^{\alpha\beta}(x), h^{\mu\nu}(y)]=-ib^{\alpha\beta\mu\nu}D(x-y)
\eqno(2.2)$$
with
$$b^{\alpha\beta\mu\nu}=\eh(\eta^{\alpha\mu}\eta^{\beta\nu}+\eta^
{\alpha\nu}\eta^{\beta\mu}-\eta^{\alpha\beta}\eta^{\mu\nu}),\eqno(2.3)$$
$$\{u^\mu(x),\tilde u^\nu(y)\}=i\eta^{\mu\nu}D(x-y)\eqno(2.4)$$
and zero otherwise. Here, $D(x)$ is the Jordan-Pauli distribution 
with mass zero and
$\eta^{\mu\nu}={\rm diag}(1,-1,-1,-1)$ the Minkowski tensor.

The gauge charge operator $Q$ is well defined through the relations
$$Q\Omega=0\eqno(2.5)$$
where $\Omega$ is the Fock vacuum and
$$d_Q h^{\mu\nu}= [Q,h^{\mu\nu}]=-{i\over 2}(\d^\nu u^\mu+\d^\mu u^\nu
-\eta^{\mu\nu}\d_\al u^\al) \eqno(2.6)$$
$$d_Q u^\mu\=d \{Q,u\}=0$$
$$d_Q\tilde u^\mu\=d \{Q,\tilde u^\mu\} =i\d_\nu h^{\mu\nu}.
\eqno(2.7)$$

The descent procedure starts from $T^{\al\be\ga}$ which must have ghost
number 3 and is totally antisymmetric. In lowest order of perturbation
theory we consider trilinear couplings only, so that we must have 3
ghost fields $u$. To exclude trivial couplings we require that
$T^{\al\be\ga}$ does not contain a co-boundary $d_QB$ for some $B\ne 0$.
Therefore we first analyse the structure of co-boundaries depending on
$u$ and derivatives of it.

It is easy to prove the following identity
$$\d_\mu\d_\nu u_\ro=id_Q\Bigl[\d_\mu h_{\nu\ro}+\d_\nu h_{\mu\ro}
-\d_\ro h_{\mu\nu}-{1\over 2}(\eta_{\nu\ro}\d_\mu h+\eta_{\mu\ro}
\d_\nu h-\eta_{\mu\nu}\d_\ro h)\Bigl],\eqno(2.8)$$
where $h=h^\mu_\mu$ denotes the trace always. This implies
$$\d_{\mu_1}\ldots\d_{\mu_p}u^\nu=d_Q b^\nu_{\mu_1\ldots\mu_p},
\quad p\ge 2\eqno(2.9)$$
for some expression $b^\nu_{\mu_1\ldots\mu_p}$; in other words all
partial derivatives of $u$ of order greater or equal to 2 are
co-boundaries. Wick products of $u$'s with one $d_QB$ are also
co-boundaries because $d_Qu=0$. This excludes higher derivative
couplings which are
sometimes considered in gravity theories. Since $T^{\al\be\ga}$
has 3 factors $u^\mu$ there can only be zero or two first derivatives of
$u$. The first possibility gives no solution in the massless case [3]
but it will appear in the massive one in the next section.

The differentiated ghost fields still contain co-boundaries due to the
identity
$$\d_\al u_\be=(\d_\al u_\be-\d_\be u_\al)+id_Q\Bigl(h_{\al\be}-
{1\over 2}\eta_{\al\be}h\Bigl).\eqno(2.10)$$
Only the antisymmetric derivatives
$$u_{\al\be}\equiv {1\over 2}\d_\al u_\be-\d_\be u_\al\eqno(2.11)$$
are not co-boundaries and should be used in $T^{\al\be\ga}$. Then there
remain the following two possibilities only:
$$u_{\al\mu}u_{\be\mu}u_\ga\quad u_{\al\be}u_{\ga\mu}u_\mu,\eqno(2.12)$$
which have to be antisymmetrized in $\al,\be,\ga$. In the second
expression the $u$ without derivative has the index $\mu$ which is
summed over, in contrast to the first expression. All products are
normally ordered.

Having determined the generic form of $T^{\al\be\ga}$ in terms of
antisymmetric derivatives we can return to ordinary partial derivatives
in the further computations due to the identity (2.10). For better
identification of the terms we shall use the convention of [3] and write
for every derivative $\d_\mu$ or $\d^\mu$ the index $\mu$ downstair, all
other Lorentz indices are written upstairs and the repeated ones are
contracted with the Minkowski tensor $\eta^{\mu\nu}$ of course. Then we
start the descent procedure with
$$T^{\al\be\ga}=a_1(u^\al_\be u^\mu u^\ga_\mu-u^\be_\al u^\mu u^\ga_\mu
-u^\ga_\be u^\mu u^\al_\mu$$
$$-u^\al_\ga u^\mu u^\be_\mu+u^\ga_\al u^\mu u^\be_\mu+
u^\be_\ga u^\mu u^\al_\mu)$$
$$+a_2(u^\mu_\al u^\be_\mu u^\ga-u^\mu_\be u^\al_\mu u^\ga
-u^\mu_\ga u^\be_\mu u^\al$$
$$-u^\mu_\al u^\ga_\mu u^\be+u^\mu_\ga u^\al_\mu u^\be+
u^\mu_\be u^\ga_\mu u^\al).\eqno(2.12)$$

Next we have to compute $\d_\ga T^{\al\be\ga}$ and this is equal to
$-id_Q T^{\al\be}$ by (1.3). To determine $T^{\al\be}$ requires an
"integration" $d_Q^{-1}$. As always in calculus this integration can be
achieved by making a suitable ansatz for $T^{\al\be}$ and fixing the
free parameters. The following 5 parameter expression will do:
$$T^{\al\be}=b_1u^\mu u^\nu_\mu h^{\al\nu}_\be+b_2u^\mu u^\al_\nu
h^{\be\nu}_\mu+b_3u^\al u^\mu_\nu h^{\be\nu}_\mu$$
$$+{b_4\over 2}u^\al_\mu u^\be_\nu h^{\mu\nu}+b_5u^\mu_\mu u^\al_\nu
h^{\be\nu}-(\al\leftrightarrow\be).\eqno(2.13)$$
Substituting this into (1.3) leads to
$$b_1=-2a_1,\> b_2=-2a_2=-2a_1,\> b_3=2a_1,\> b_4=-4a_1,\> b_5=-2a_1.$$  
An overall factor is arbitrary, we take $a_1=-1$ which gives
$$T^{\al\be}=2(u^\mu u^\nu_\mu h^{\al\nu}_\be+u^\mu u^\al_\nu
h^{\be\nu}_\mu-u^\al u^\mu_\nu h^{\be\nu}_\mu
+u^\al_\mu u^\be_\nu h^{\mu\nu}+u^\nu_\nu u^\al_\mu h^{\be\mu})
-(\al\leftrightarrow\be).\eqno(2.14)$$

Next in the same way we compute $\d_\be T^{\al\be}$ and make an ansatz 
for $T^\al$.
The latter now has to contain ghost-antighost couplings also. The
precise form can be taken from the following final result:
$$T^\al=4u^\mu h^{\be\nu}_\mu h^{\al\nu}_\be-2u^\mu h^{\be\nu}_\mu
h^{\be\nu}_\al-2u^\al h^{\mu\nu}_\be h^{\be\nu}_\mu
-4u^\be_\nu h^{\al\be}_\mu h^{\mu\nu}+4u^\nu_\nu h^{\al\be}_\mu h^{\be\mu}
+u^\al h^{\mu\nu}_\be h^{\mu\nu}_\be- $$
$$-2u^\nu_\nu h^{\mu\be} h^{\mu\be}_\al-{1\over 2}u^\al h_\mu h_\mu
+u^\nu_\nu hh_\al+u^\nu h_\nu h_\al
-2u^\mu_\nu h^{\mu\nu}h_\al+4u^\mu_\nu h^{\mu\be}_\al h^{\be\nu}
-4u^\mu_\nu h^{\al\be}_\mu h^{\be\nu}$$
$$-2u^\mu u^\nu_\mu\tilde u^\nu_\al+2u^\mu u^\al_\nu\tilde u^\nu_\mu
-2u^\al u^\mu_\nu\tilde u^\nu_\mu+2u^\nu_\nu u^\al_\mu\tilde u^\mu
+2u^\mu u^\nu_{\nu\mu}\tilde u^\al-2u^\al u^\mu_{\mu\nu}\tilde u^\nu.
\eqno(2.15)$$

The last step calculating $\d_\al T^\al$ and setting it equal to
$-id_Q T$ gives the trilinear coupling of massless gravity
$$T=-h^{\al\be}h_\al h_\be+2h^{\al\be}h^{\mu\nu}_\al h^{\mu\nu}_\be
+4h^{\al\be}h^{\be\mu}_\nu h^{\al\nu}_\mu
+2h^{\al\be}h^{\al\be}_\mu h_\mu-4h^{\al\be}h^{\al\mu}_\nu h^{\be\mu}_\nu$$
$$-4u^\mu\tilde u^\nu_\be h^{\nu\be}_\mu+4u^\be_\nu\tilde u^\be_\mu 
h^{\mu\nu}-4u^\nu_\nu\tilde u^\be_\mu h^{\be\mu}+4u^\mu_\nu\tilde u^\be_\mu
h^{\nu\be}.\eqno(2.16)$$
This is in agreement with eq.(5.7.1) of [3] apart from a small change
in the ghost couplings. As shown in [3] sect.5.5 the pure graviton part
agrees with the expansion of the Einstein-Hilbert Lagrangian.

\section{Massive gravity}

Now we assume that all fields are massive and satisfy the Klein-Gordon
equation with mass $m$:
$$(\sq+m^2)h^{\mu\nu}=0=(\sq+m^2)u^\mu\ldots    \eqno(3.1)$$
In the massive case in addition to the previous fields we need a vector
field $v^\mu(x)$ ($(\sq+m^2)v^\mu=0$) called vector-graviton field for
some reason [5]. These fields are quantized as follows [1]
$$[h^{\alpha\beta}(x), h^{\mu\nu}(y)]=-ib^{\alpha\beta\mu\nu}D_{m}(x-y)
\eqno(3.2)$$
$$\{u^\mu(x),\tilde u^\nu(y)\}=i\eta^{\mu\nu}D_{m}(x-y)\eqno(3.3)$$
$$[v^\mu(x),\, v^\nu(y)]={i\over 2}\eta^{\mu\nu}D_{m}(x-y),$$ 
$$d_Q u^\mu\=d \{Q,u\}=0$$
remain unchanged, but
$$d_Q\tilde u^\mu\=d \{Q,\tilde u^\mu\} =i(\d_\nu h^{\mu\nu}-m v^\mu)
\eqno(3.6)$$
$$d_Q v^\mu\=d [Q, v^\mu]=-{i\over 2}mu^\mu\eqno(3.7)$$
are new.

>From (3.7) it follows that $u^\al=d_Qv^\al 2i/m$ is a co-boundary. Therefore
the cohomological argument at the beginning of the last section breaks
down. Instead we require for physical reasons that the massive theory
has a smooth limit for $m\to 0$. Then the previous expression (2.12) for
$T^{\al\be\ga}$ must only be modified by a simple mass term
$$T^{\al\be\ga}_m=T^{\al\be\ga}+am^2u^\al u^\be u^\ga.\eqno(3.8)$$
This mass term is needed to compensate mass terms from the Klein-Gordon
equation when we compute
$$\d_\ga T^{\al\be\ga}_m=(\d_\ga T^{\al\be\ga})_0-m^2u^\al u^\mu
u^\be_\mu+m^2u^\be u^\mu u^\al_\mu$$
$$-m^2u^\mu u^\be_\mu u^\al+m^2u^\mu u^\al_\mu u^\be$$
$$+am^2(u^\al_\ga u^\be u^\ga+u^\al u^\be_\ga u^\ga+u^\al u^\be u^\ga_\ga).
\eqno(3.9.)$$
Here the subscript 0 means always the exact zero mass expression from
sect.2. The additional mass terms come from those terms in (2.12) which
have a derivative index $\ga$ downstairs such that with the $\d_\ga$ in
(3.9) a wave operator is obtained. Now
most of the mass terms in (3.9) cancel if we choose $a=-2$.

The last term $u^\al u^\be u_\ga^\ga$ in (3.9) survives, it gives rise 
to a deformation of $T^{\al\be}$ (2.14):
$$T^{\al\be}_m=T^{\al\be}+2m(u^\be u^\mu_\mu v^\al-u^\al u^\mu_\mu v^\be).
\eqno(3.10)$$
Indeed, $-id_Q$ of the additional mass terms cancels the last mass terms
in (3.9). In the same manner we compute
$$\d_\be T^{\al\be}_m=(\d_\be T^{\al\be})_0-2m^2u^\mu u^\nu_\mu 
h^{\nu\al}$$
$$+2m(u^\be_\be u^\mu_\mu v^\al+u^\be u^\mu_{\mu\be} v^\al+
u^\be u^\mu_\mu v^\al_\be-u^\al_\be u^\mu_\mu v^\be-u^\al u^\mu_{\mu\be}
v^\be-u^\al u^\mu_{\mu}v^\be_\be),\eqno(3.11)$$
where the first term in the bracket vanishes. Calculating $-id_QT^\al$
from the massles expression (2.15) we now get additional terms $mv^\nu$
from $-id_Q\tilde u^\nu$. Consequently, (3.11) is equal to
$$-id_QT^\al+2m(u^\mu u^\nu_\mu v^\nu_\al-u^\mu u^\al_\nu v^\nu_\mu
+u^\al u^\mu_\nu v^\nu_\mu-u^\nu_\nu u^\al_\mu v^\mu
-u^\mu u^\nu_{\nu\mu} v^\al+u^\al u^\mu_{\mu\nu} v^\nu)$$
$$+2m(u^\be u^\mu_{\mu\be}v^\al+u^\be u^\mu_{\mu}v^\al_\be-
u^\al_\be u^\mu_{\mu}v^\be-u^\al u^\mu_{\mu\be}v^\be-u^\al u^\mu_{\mu}v^\be 
_\be)$$
$$-2m^2u^\mu u^\nu_{\nu\mu}h^{\nu\al}.$$
$$=-id_QT^\al+2mu^\mu(u^\nu_\mu v^\nu_\al-u^\al_\nu v^\nu_\mu+
u^\be_\be v^\al_\be)$$
$$+2mu^\al(u^\mu_\nu v^\nu_\mu-u^\mu_\mu v^\be_\be)-2m^2u^\mu 
u^\nu_{\nu\mu}h^{\nu\al}.\eqno(3.12)$$
This implies
$$T^\al_m=T^\al+4u^\mu v^\nu_\mu v^\nu_\al-2u^\al v^\nu_\mu v^\nu_\mu$$
$$+4m(u^\al v^\nu_\mu h^{\mu\nu}-u^\mu v^\nu_\mu h^{\al\nu})-
m^2u^\al(h^{\mu\nu}h^{\mu\nu}-{1\over 2}h^2).\eqno(3.13)$$

In the last step we have
$$\d_\al T^\al_m=(\d_\al T^\al)_0+m^2u^\nu_\nu h^{\mu\be}h^{\mu\be}
-{m^2\over 2}u^\nu_\nu h^2+2m^2u^\mu_\nu h^{\mu\nu}h$$
$$-4m^2u^\mu_\nu h^{\mu\be}h^{\nu\be}+2m^2u^\mu u^\nu_\mu\tilde u^\nu$$
$$+4u^\mu_\al v^\nu_\al v^\nu_\mu-4m^2u^\mu v^\nu_\mu v^\nu-
2u^\al_\al v^\nu_\mu v^\nu_\mu$$
$$+4m(u^\al_\al v^\nu_\mu h^{\mu\nu}+u^\al v^\nu_\mu h^{\mu\nu}_\al
-u^\mu_\al v^\nu_\mu h^{\al\nu}-u^\mu v^\nu_\mu h^{\al\nu}_\al).$$
where the first $m^2$-terms come from the Klein-Gordon equation.
This is equal to
$$=-i(d_QT)_0+4m(u^\mu v^\nu_\be h^{\nu\be}_\mu-u^\be_\nu v^\be_\mu
h^{\mu\nu}+u^\nu_\nu v^\be_\mu h^{\mu\be}-u^\mu_\nu v^\be_\mu h^{\nu\be})$$
$$+m^2d_Q\Bigl({4\over 3}h^{\mu\nu}h^{\mu\be}h^{\nu\be}- 
h^{\mu\be}h^{\mu\be}h+{1\over 6}h^3\Bigl)$$
$$+4md_Q(v^\nu_\mu u^\mu\tilde u^\nu)-4d_Q(h^{\mu\al}v^\nu_\al
v^\nu_\mu).\eqno(3.14)$$
Now we can read off the coupling of massive gravity:
$$T_m=T+m^2\Bigl({4\over 3}h^{\mu\nu}h^{\mu\be}h^{\nu\be}-
h^{\mu\be}h^{\mu\be}h+{1\over 6}h^3\Bigl)$$
$$+4m u^\mu\tilde u^\nu v^\nu_\mu-4h^{\mu\al}v^\nu_\al v^\nu_\mu.
\eqno(3.15)$$

This result (3.15) agrees with our previous one (2.9) in ref.[1]
up to a different ghost coupling in $T$. It is simple to see that this
difference of the two ghost couplings is a divergence and, therefore,
uninteresting. The $T^\al$ from [1] (eq.(2.10)) differs from 
(3.13) also in a trivial way by $d_Q D^\al+\d_\be D^{\al\be}$
with $D^{\al\be}$ antisymmetric. These differences, however,
lead to much more complicated descent equations.

Gauge invariance of second order can be verified as in [1]. Alternatilvely, one can
try to prove that if one modifies the Lagrangian by a total derivative and/or a coboundary
then the higher-order chronologial products are also trivially modified; such a result is valid
for Yang-Mills theories [3][6]. A gravitational Higgs field is not needed in the massive spin-2 case
in contrast to spin-1 [3]. As shown in [1] the second order couplings
correspond to non-linear classical gravity in the following sense:
Consider Einstein's Lagrangian with a cosmological constant $\Lambda$
$$L_E=-{2\over\kappa^2}\sqrt{-g}(R-2\Lambda),\quad
\kappa^2=32\pi G,\eqno(3.16)$$
where $g={\rm det}(g_{\mu\nu})$ and $G$ is Newton's constant. For the
expansion in powers of $\kappa$ it is convenient to use the so-called
Goldberg variables
$$\tilde g^{\mu\nu}=\sqrt{-g}g^{\mu\nu}=\eta^{\mu\nu}+\kappa h^{\mu\nu}.
\eqno(3.17)$$
Then the quadratic terms $O(\kappa^0)$ determine the free theory.
The corresponding Euler-Lagrange equation leads to the Klein-Gordon
equation with mass $m^2=2\Lambda$. That means the inverse graviton 
mass must be of the order of the current Hubble scale. The cubic
part $O(\kappa^1)$ agrees with the $h$-terms in (3.15) and the
quartic part with the second order couplings derived from gauge
invariance. This shows that the massive spin-2 gauge theory corresponds
to Einstein gravity with a cosmological term. As discussed in [5]
the massive graviton has 6 physical degrees of freedom. Nevertheless,
there is no conflict with observations of gravitational radiation:
the cross sections for bremsstrahlung in the massive case differ
from the massless ones only by a tiny correction of order $O(m)$ [5].

\end{document}